\documentstyle[11pt,aaspp4]{article}

\begin{document}

\received{}
\revised{}
\accepted{}

\lefthead{Catelan \& de Freitas Pacheco}
\righthead{Clump morphology in metal-rich globulars}

\slugcomment{To appear in PASP (February issue)}

\singlespace

\title{Metal-rich globular clusters of the Galaxy:
       morphology of the helium-burning ``clump," and the determination
       of relative ages through the ``$\Delta V$" method}
\author{M. Catelan}
\affil{Instituto Astron\^omico e Geof\'\i sico, Universidade de S\~ao Paulo\\
       C.P. 9638, CEP 01065-970 S\~ao Paulo, SP\\
       Brasil\\
       e-mail: marcio@vax.iagusp.usp.br}
\and
\author{J. A. de Freitas Pacheco}
\affil{Observatoire de la C\^ote d'Azur\\
       B.P. 229, 06304 Nice Cedex 04\\
       France\\
       e-mail: pacheco@rameau.obs-nice.fr}
\begin{abstract}
Morphological aspects of the $V,\, \bv$ diagram of metal-rich globular
clusters are analyzed, on the basis of stellar evolution models for
horizontal-branch (HB) and red giant branch stars. These have been incorporated
into detailed synthetic HB models, the influence of differential reddening
upon which is also discussed. The synthetic HB models are found to be very
clumpy in the colour-magnitude diagram, though more extended structures
occasionally result due to the effect of evolution away from the zero-age HB.
In particular, ``sloped" clumps may be naturally expected for sufficiently
high helium abundances. Differential reddening is found to have a smaller
influence than evolution itself upon the morphology of the clump.
The synthetic HB models are also used to study the age difference between
metal-rich clusters on the basis of the ``$\Delta V$" method.
We emphasize the importance of knowing in advance the relative abundances
of helium and metals in order to estimate this age difference quantitatively.
We present chemical evolution models that serve to illustrate the effect, and
show that, for two clusters of identical [Fe/H] and $\Delta V$, the one
enriched
predominantly by ejecta from supernovae type II explosions will appear
{\em younger} than the other which also suffered enrichment from supernovae
type Ia ejecta. The implied locations of the red giant branch feature known
as the ``bump" are also discussed on the basis of recent observational and
theoretical results. Our corresponding predictions should be checked against
accurate observational data obtained with the {\em Hubble Space Telescope}.
\end{abstract}


\clearpage

\section{Introduction}

Recently, CCD photometry of metal-rich Galactic globular clusters (MRGCs),
such as NGC 6553 (Ortolani, Barbuy \& Bica 1990, hereinafter referred to as
OBB90) and Terzan 1 (Ortolani, Bica \& Barbuy 1993), has been obtained with
the purpose of revealing the morphology of the main branches in their
colour-magnitude diagrams (CMDs), and, when possible, to investigate their
ages and establish a chronology for the formation of the different structures
(halo, disk, bulge) that comprise the present-day Galaxy. Many interesting
results emerged from these investigations, among which the suggestion
of a possible ``tilted" helium-burning ``clump" or horizontal branch (HB) in
the $V,\, \bv$ CMDs of MRGCs (OBB90; see also Armandroff 1988). However, the
significance of this result was not well established in these studies, and
indeed some controversy still exists as to its reality and extension.
{}From the theoretical point of view, in particular, the present authors are
aware of no investigation aiming at a more detailed analysis of the allowed
clump morphologies in MRGCs on the basis of available evolutionary tracks,
except for the study of Dorman, VandenBerg \& Laskarides (1989) on 47 Tuc.
In addition, in spite of the remarkable observational efforts by the above
and other authors, no conclusive results could be obtained about the ages
of the studied objects --- especially those located in the Galactic bulge.

Establishing a reliable chronology for the formation of the Galactic subsystems
will be possible in the near future. Indeed, several observational projects
on many MRGCs that are presently being carried out by different groups
employing the {\em Hubble Space Telescope} ({\em HST}) will soon be providing
us with CMDs of unprecedented accuracy, and standard age-determination
techniques will be able to shed light on the problem of the relative ages
of the halo, bulge, and disk.

In the present study, we carry out a theoretical investigation of the
morphology of the HB clump in MRGCs on the basis of the detailed synthetic
HB models which are described in Sect. 2. We demonstrate, in Sect. 3, that
the production of synthetic models with tilted clumps is not a {\em general}
consequence of evolution away from the zero-age HB (ZAHB), even though it can
be achieved for models with sufficiently high helium abundances, and/or in
fields where differential reddening is very large. On the basis of the
``$\Delta V$" method and different chemical evolution scenarios for the
MRGCs of given $\Delta V$ (the magnitude difference between the clump and
the turnoff), the theoretical problem of the determination of the age
difference between two MRGCs is investigated in Sect. 4. We show that
knowledge of the {\em relative} abundances of helium and metals is essential
in order to determine the difference in age between any two MRGCs in a
reliable way. In particular, we emphasize that this may be a most important
point when comparing bulge and disk globulars of similar [Fe/H], since their
chemical evolution histories may have been different. For this reason, a strong
effort must be made to determine the abundances of helium and the
$\alpha$-capture elements in these clusters {\em from the observations}.
In Sect. 5, we evaluate the theoretically-expected location of the
red giant branch (RGB) feature known as the ``bump" with respect to the clump.
Finally, in Sect. 6 we summarize our results and provide some additional
discussion.

\section{Synthetic models of the HB clump}

Catelan (1993) has computed synthetic HB models for low metallicities in the
theoretical plane. In the present work, we extend his calculations by including
published evolutionary tracks for higher metallicities. The evolutionary masses
considered match the HB morphology that is usually observed in metal-rich
objects, where the stars lie entirely to the red of the instability strip
(Armandroff 1988).

The incorporated evolutionary tracks for HB stars include the models computed
by Sweigart (1987, hereinafter SW87) for $(Y_{{\rm MS}},\,Z) = (0.25,\,0.01)$
and $(Y_{{\rm MS}},\,Z) = (0.30,\,0.01)$, as well as those by Castellani,
Chieffi \& Pulone (1991, hereinafter CCP91) --- based upon improved input
physics ---  for $(Y_{{\rm MS}},\,Z) = (0.23,\,0.006)$ and
$(Y_{{\rm MS}},\,Z) = (0.23,\,0.02)$. $Y_{{\rm MS}}$ represents, as usual,
the helium abundance prior to the main-sequence phase.

To perform the transposition of the models from the theoretical to
the observational ($M_V,\, \bv$) plane, we employ the conversions from
VandenBerg 1992 (hereafter VdB92), which are actually a smooth match of
the VandenBerg \& Bell (1985) atmospheres to those of Bell \& Gustafsson
(1978) and of Kurucz (1979). His published tables are restricted to the range
$3500\,{\rm K}\,< T_{{\rm eff}} < 8000\,{\rm K}$ --- which turns out to be
adequate for our present purposes. It is important to note that the specific
procedures adopted by VdB92 in obtaining these conversions do not introduce
significant uncertainties into our results, since both the colour
transformations and bolometric corrections $BC_V$ of VdB92, in the
temperature and gravity range of our models (cf. Figs. 1 -- 4), coincide
closely with those from Bell \& Gustafsson 1978. As recommended by VdB92,
Hill's (1982) Hermite interpolation algorithm has been employed for
interpolation with respect to $T_{{\rm eff}}$ and $\log g$, and linear
interpolation with respect to [M/H].

Observational errors in the photometry have also been included in the present
models. We adopt $\sigma_{V} = 0.035$ mag and $\sigma_{B-V} = 0.046$ mag at
the level of the HB, as estimated from the study of OBB90 for the MRGC NGC
6553.

We plan to extend the present calculations in the future by including the
transformations to the infrared planes, employing the prescriptions of
Bell \& Gustafsson (1989) and Bell (1992).

\section{Expected morphology of the metal-rich clumps}

Figs. 1, 2 and 3 demonstrate that the production of tilted HB clumps is
not generally expected, but rather a matter of fine-tuning of evolutionary
parameters of HB evolution. Figs. 1a and b demonstrate that the CCP91
evolutionary tracks do not produce significantly tilted synthetic HB models,
irrespective of $Z$. This is probably related to the low helium abundance
employed in the CCP91 calculations: the SW87 tracks for
$(Y_{{\rm MS}},\,Z) = (0.25,\,0.01)$ do not lead to sloped models either (cf.
Figs. 1c and 2a), as opposed to the $(Y_{{\rm MS}},\,Z) = (0.30,\,0.01)$
combination (cf. Fig. 3a). Indeed, in the latter case, for
$\langle M_{{\rm HB}}\rangle \simeq 0.66 - 0.68\, M_{\sun}$,
tilted models are clearly obtained.

Inspection of Figs. 7 through 9 in SW87 does show that, for stars in this mass
range, the ``blueward loops" (which characterize the HB evolutionary tracks
for stars in which the H-burning shell is a major contributor to the total
luminosity; cf. Iben \& Rood 1970) present a trend of increasing slope
${\rm d}\log\,L/{\rm d}\log\,T_{{\rm eff}}$ with increasing $Y$. A similar
conclusion may be drawn from close inspection of the diagrams published by
Seidel, Demarque \& Weinberg (1987), but the effect is especially clear in
Figs. 1 through 5 of Sweigart \& Gross (1976), where the computations were
extended up to $Y = 0.40$, and in the diagrams presented by Dorman et
al.\ (1989) in their detailed analysis of the HB morphology in 47 Tuc.
Sweigart \& Gross interpret this increase in slope of the blueward loops
with increasing $Y$ as being a direct consequence of the increasing rate
of shell H-burning, which makes the H-shell advance outward more rapidly
through the stellar envelope. Thus, we expect that a very high helium
abundance for the bulge stars, as might be present if a large contribution
to the chemical enrichment from type II supernovae took place (cf. Sect. 4
below), would lead more naturally to sloped HBs. Unfortunately, we cannot
present synthetic models for $Y_{\rm MS} > 0.30$, since the required
extensive grids of HB evolutionary tracks are presently lacking.

This result, in fact, is not new: Dorman et al.\ (1989) have suggested
employing the slope of the red clump as a diagnostic of the helium abundance
in GCs with red HBs, and from this have estimated that $Y$ in 47 Tuc probably
lies midway between $0.20$ and $0.30$.

The reader will find instructive to compare the synthetic model in Fig. 1a
with the observed CMDs for 47 Tuc that have been published by Hesser et al.\
(1987) and Carney, Storm \& Williams (1993). The general agreement appears
quite satisfactory. Note that the final evolution to the asymptotic GB, which
is clear in the observed CMD, is present in the CCP91 tracks, but lacking in
the SW87 case. Note as well that the slope of the lower envelope of the HB
distribution, which is present in the synthetic model {\em but not in ZAHB
sequences}, serves as a major caution against using ZAHB models in
applications to clusters with red HBs (Dorman et al.\ 1989).

The presence of a tilt in the $M_V, \, \bv$ diagram in the cases where there
was none in the corresponding theoretical diagram is a consequence of the
blanketing contained in the VdB92 transformations, which affects stars in
a different way as a function of their temperatures along the HB (compare
Figs. 3a and 4). Indeed, inspection of Fig. 7 of VdB92 shows that, in the
relevant temperature range,
$3 \lesssim {\rm d}\, BC_V/{\rm d}\log\,T_{{\rm eff}} \lesssim 6$. If such a
trend is not counterbalanced by an opposite slope of the blueward loop
in the $M_{\rm bol} - \log\,T_{\rm eff}$ diagram (cf. Fig. 4), a tilted model
will clearly result --- which is the reason for the peculiar slope of the model
in Fig. 3a. This effect is also clear in Dorman's (1992) Figs. 10 and 11.

Differential reddening is generally expected to affect the CMD morphology
of most MRGCs, since they generally lie very close to the Galactic plane.
According to OBB90, for instance, in the specific case of NGC 6553 this
effect may account for a vertical dispersion of $\Delta V \simeq 0.18$ mag
and a horizontal dispersion of $\Delta (\bv )\simeq 0.06$ mag in the
cluster's CMD. Thus, differential reddening works in the sense of bringing
the clumpy models (cf. Figs. 1 and 2a) into tilted structures.

This is confirmed by our synthetic HB models. The effect is illustrated in
Figs. 2b--d and 3b. In the former panels, increasing amounts of (uniform)
differential reddening are assumed in the model construction, in comparison
with the unreddened diagram depicted in Fig. 2a. Similarly, in Fig. 3b the
unreddened synthetic HB displayed in Fig. 3a is changed in order to simulate
the effects of the differential reddening that OBB90 estimate to be present
in the field of NGC 6553 [i.e., $\Delta E(\bv ) \simeq 0.06$ mag].
Differential ``reddening," in these simulations, is defined so that the mean
reddening amounts to zero.

It is clear, from Figs. 2a--b and 3a--b, that a differential reddening as
small as $\Delta E(\bv ) = 0.06\,\,{\rm mag}$ {\it cannot cause a CMD
dispersion as large as the one originating from the evolution away from
the ZAHB itself.} Thus, unless differential reddening in the field of a
MRGC is very large (Figs. 2c--d), it seems unlikely that sloped HBs can
be produced on this basis alone.

Thus, in the case of the MRGC NGC 6539, it is very likely that the primary
cause for the sloping HB is indeed the large amount of differential reddening
that is present in the field of this cluster, as Fig. 7 in Armandroff 1988
convincingly demonstrates. In the case of NGC 6553, on the other hand, this
possibility may appear somewhat less appealing.

To be sure, the above conclusions rely on the adequacy of the VdB92 blanketings
for cool stars. Dorman (1994) finds that the VdB92 transformations can
reproduce
the observational data quite satisfactorily. At any rate, inclusion of the
Kurucz (1991, 1992) and Bell \& Tripicco (1995) blanketings should prove of
interest in the present context. Similarly, a more extended and homogeneous
set of updated evolutionary tracks for metal-rich compositions would definitely
prove very useful. Also, the computation of synthetic models in the infrared
would be of great interest for a variety of reasons: for instance, while MRGCs
are heavily obscured at optical wavelengths, in the infrared region they are
less subject to interstellar extinction (e.g., Guarnieri et al.\ 1995), which
also renders the effect of differential reddening a less important problem
(Davidge \& Simons 1994). Transforming models similar to the present ones to
the infrared plane on the basis of the model atmospheres by Bell \& Gustafsson
(1989) and Bell (1992) is a necessary future step in the investigation of the
expected morphology of HB clumps in MRGCs. We emphasize the need of synthetic
calculations employing {\em extensive grids} (in $Y$, $Z$, and $M_{\rm HB}$)
of HB evolutionary tracks to perform more detailed comparisons with the
observed CMDs, since neither ZAHB sequences nor individual evolutionary
tracks are expected to be able to adequately represent the observed
structures (Dorman et al.\ 1989).

Before closing the present section, we believe it may be of interest to note
that ZAHB sequences do become themselves tilted structures at sufficiently
high values of $M_{{\rm HB}}$ (e.g., Seidel et al.\ 1987). This may also
contribute to the possible tilting of HB clumps. For instance, Fig. 3 in
Tripicco, Dorman \& Bell 1993 shows that, for $Y_{{\rm MS}} = 0.27$ and
$Z = 0.02$, the sign of ${\rm d}\log\,L/{\rm d}\log\,T_{{\rm eff}}$ for the
ZAHB becomes positive for $M_{{\rm HB}} \ga 0.9\, M_{\sun}$. Our models have
all been built for significantly lower masses. However, this would demand ages
$\la\, 8.9$ Gyr for a (rather modest) overall mass loss on the RGB of
$\Delta M \sim 0.1\, M_{\sun}$. Thus, this only becomes a realistic
possibility for sufficiently young MRGCs.

\section{The age difference between MRGCs from the $\Delta${\it V} method}

\subsection{Theoretical approach}

There is not a sufficiently accurate method to estimate absolute ages of MRGCs.
For instance, the $\Delta V$ method cannot be directly employed due to the many
uncertainties involved in, e.g., the helium abundance and the HB magnitude
scale. Even relative ages may be difficult to determine. The
$\Delta (\bv )_{\rm TO,RGB}$ method, while very useful in applications to
metal-poor objects, cannot be reliably applied to MRGCs, since age and [Fe/H]
differences may be hard to disentangle at the metal-rich end (VandenBerg,
Bolte \& Stetson 1990; Bolte 1992). This, unfortunately, is a most important
point in what regards MRGCs, since [Fe/H] values for these objects are often
known only with large uncertainties (Bolte 1992).

The following discussion is aimed at highlighting the important ingredients
that even a differential age determination for MRGCs based upon the $\Delta V$
method should include.

{}From Eq. (1) in Renzini 1991 and the analysis of Chieffi, Straniero \&
Salaris
(1991; see also Salaris, Chieffi \& Straniero 1993), one derives

\begin{displaymath}
\Delta\log t_9 = 0.37\,
[\Delta M_{V}^{{\rm HB}} + \Delta (\Delta V_{{\rm HB}}^{{\rm TO}})]
- 0.43\, \Delta Y_{{\rm MS}}
- 0.13\, \Delta {\rm [Fe/H]} - 0.13\, \Delta\log(0.579\, f + 0.421),
\end{displaymath}

\noindent where $\Delta V_{{\rm HB}}^{{\rm TO}}$ is the difference in magnitude
between the turn-off region and the HB, and $f = 10^{[\alpha/{\rm Fe}]}$. We
assume this relation to be approximately valid up to near-solar metallicities.
We see that helium and the $\alpha$-elements play a r\^ole. To be noted, in
particular, is that $M_V^{\rm HB}$ depends on $Y_{\rm MS}$, [Fe/H], and $f$
as well. These abundances are not directly known for MRGCs from the
observations, and one must often resort to theoretical scenarios to estimate
them (e.g., Renzini \& Greggio 1990). From the following discussion, it will
become clear that observational determinations of these quantities are urgently
needed --- and in fact at least the [$\alpha$/Fe] ratio may be obtained
directly from spectroscopic measurements.

The time evolution of the abundance of a given element is not well predicted
yet. Among the reasons for this is the controversy that still exists about
the enrichment timescale for metal-poor stars. While the onset of type Ia
supernovae (SN Ia) may require a relatively long scale $\tau \gtrsim 10^9$ yr
(van den Bergh 1991), the evidence from kinematical studies of field stars
and the analysis of CMDs of GCs remains somewhat controversial. Since
addressing these subjects is beyond the scope of our article, we refer the
reader to the following papers for additional information and references:
Beers \& Sommer-Larsen 1995 (field-star kinematics); Catelan \& de Freitas
Pacheco 1995 (ages of the halo GCs); Marquez \& Schuster 1994 (ages of field
stars). The reviews by Majewski (1993) and Matteucci (1995) address these
points at some length.

These uncertainties notwithstanding, element ratios as a function of
metallicity can be computed, following the method by de Freitas Pacheco (1993)
and Barbuy, de Freitas Pacheco \& Castro (1994). Such a procedure requires the
knowledge of the relative yields due to massive stars (SN type II and maybe
type Ib) and to SN Ia. The relative number of events of different supernovae
types is also necessary in order to compute these chemical evolution models.

For the present purposes, we have considered three distinct scenarios which
should serve to illustrate reasonably well the associated uncertainties in
applications of the $\Delta V$ method. In the first (Case 1), we assume that
in the metallicity range $-3.0 < {\rm [Fe/H]} < +0.5$ the chemical enrichment
was produced by SN II only. In the second (Case 2), we consider that at
${\rm [Fe/H]} = -1.0$ occurs the onset of SN Ia, with an increasing
contribution to the yields until a maximum of 10\% in the relative number of
events. In the last scenario (Case 3), we examine a higher contribution of
SN Ia to the chemical enrichment, amounting to about 40\% in the relative
number at ${\rm [Fe/H]} = +0.5$.

The yields from massive stars were computed by averaging over the initial mass
function (Salpeter's 1955 law) the results by Arnett (1990). For SN Ia,
we used the yields of model W7 by Nomoto, Thielemann \& Yokoi (1984). The
initial He abundance (cosmological value) was taken to be equal to 0.075 by
number. Fig. 5a shows the He abundance by mass as a function of the metallicity
[Fe/H]. Fig. 5b shows the [O/Fe] ratio (representative of the abundance of the
$\alpha$-elements) versus [Fe/H] for the three considered scenarios.

Let us examine the age difference between two given GCs. For MRGC1, we assume
Case 1 models to be valid, and adopt ${\rm [Fe/H]} = -0.71$. Hence,
$Y_{{\rm MS}}\simeq 0.266$ and [$\alpha$/Fe] $\simeq +0.35$ for this object
(cf. Fig. 5a). For MRGC2, in turn, we investigate the implications of all
three different chemical evolution scenarios, assuming ${\rm [Fe/H]} = -0.2$.
The resulting abundance parameters are displayed in Table 1.

According to Eq. (1), an estimate of the difference in {\em absolute magnitude}
between the HBs of the two MRGCs is also necessary. This was accomplished as
follows. On the basis of the discussion in the next section, we adopt
$\Delta M_{V}^{{\rm HB}}/\Delta\log\,Z \approx (1.33 - 0.98)/(-1.7 + 2.22)
\simeq 0.67$.
Following our synthetic clump models based upon the SW87 tracks for
$Y_{{\rm MS}} = 0.25$ and $Y_{{\rm MS}} = 0.30$, we take
$\Delta M_{V}^{{\rm HB}}/\Delta Y_{{\rm MS}} \approx (0.6 - 0.8)/(0.30 - 0.25)
= -4$. The resulting $\Delta M_{V}^{{\rm HB}}$ values (MRGC1 minus MRGC2)
are also displayed in Table 1, as are the inferred age differences.

[It is to be noted, from these estimates, that a ratio of $\approx -6$ is
found between the $Y_{{\rm MS}}$ and $\log\,Z$ dependencies of $M_V^{{\rm HB}}$
from the models. On the other hand, from Fig. 1 in Demarque \& Lee 1992 a
ratio of $\approx -10$ is estimated instead. Since the latter models are also
based upon older input physics in comparison with the CCP91 study, this
clearly illustrates the effect of improved opacities upon the model
predictions for high-metallicity stars.]

In Table 1, a value of $\Delta V = 3.6$ has been assumed for MRGC1, which
is a quite normal value found among old halo GCs. For MRGC2, two different
values have been considered, $\Delta V = 3.3$ and $3.6$. In fact, Fig. 5
and Eq. (1) may easily be employed for the determination of age differences
for arbitrary [Fe/H] and $\Delta V$.

We see, from an analysis of Table 1, that {\em the different assumptions about
the chemical evolution of the metal-rich system affect the derived age
differences in a significant way}. Note, in particular, that for a given
$\Delta V$ value, MRGCs which have been formed from gas enriched primarily
by SN II explosions appear to be somewhat {\em younger\/} than those which
have also been subject to SN Ia ejecta (Cases 2 and 3). This may be a most
important point when comparing ages of disk and bulge globulars of similar
[Fe/H].

To illustrate the point, we call the reader's attention to the following
quantitative estimates. MRGC1 will be considered older by $15 \%$ than
MRGC2 for $\Delta V({\rm MRGC2}) = 3.60,\, 3.45$, and 3.33 in Cases 1, 2, and
3, respectively. Similarly, coevality between the two objects will be reached
for $\Delta V({\rm MRGC2}) = 3.77,\, 3.61,\, 3.49$ in these three Cases,
respectively. As before, ${\rm [Fe/H]} = -0.2$ has been assumed for MRGC2,
and ${\rm [Fe/H]} = -0.71$ and $\Delta V = 3.6$ for MRGC1.

The above conclusions would change if, as pointed out at the end of Sect. 3,
the clump structure is populated by stars more massive than
$\approx 1.0\, M_{\sun}$. In this case, the HB of MRGC2 (which has a higher
[Fe/H] value) could be brighter by $\approx 0.35$ mag than assumed herein
(Castellani, Chieffi \& Straniero 1992). As a consequence, as is apparent
from Eq. (1), the age difference would increase (i.e., MRGC2 would be younger),
in agreement with the requirement of smaller ages to account for the higher
masses that would be reached at the tip of the RGB.

\subsection{On applications of the method to real clusters}

In the first place, one should note that the above results have been inferred
on the basis of a theoretical analysis of the variation of the HB magnitude
with the metallicity and helium abundance. Observational estimates of the
magnitudes of the core helium-burning stars are usually obtained for more
metal-poor GCs with populated instability strips (see, e.g., Sandage \&
Cacciari 1990; Sandage 1993; Carney, Storm \& Jones 1992; for recent
discussions
and references), which is obviously not the case here. Moreover, applying
relations obtained for the RR Lyrae level to objects lying to the blue or
to the red of the instability strip is not a straightforward procedure:
theoretically-estimated corrections are needed with the purpose of roughly
obtaining the magnitude of the clump of red HB stars from the estimated
magnitudes of the (often missing) RR Lyrae variables (cf. Catelan 1992, 1993
and references therein). For this reason, a comparison with relations obtained
for samples of field RR Lyrae stars which do contain a few metal-rich variables
(e.g., Skillen et al.\ 1993) should also be made with caution. In addition,
much controversy still exists in what concerns the slope of the
$M_V^{{\rm HB}}-{\rm [Fe/H]}$ relation. This question has been addressed
quite recently by several authors (e.g., Carney et al. 1992; Catelan 1992;
Fernley 1993; Sandage 1993), without a general consensus having been found.

Are the chemical abundances predicted by the above chemical evolution
models representative of real GCs? As to the model parameters for MRGC1,
we note that they may characterize rather satisfactorily the well-studied
GC 47 Tuc. This cluster, according to Zinn (1985), has ${\rm [Fe/H]} = -0.71$.
The helium abundance implied by our Case 1 models is not incompatible with
estimates based upon the ``$R$-method" (Buzzoni et al.\ 1983; Caputo,
Martinez Roger \& Paez 1987; Hesser et al.\ 1987; CCP91), while the abundances
of the $\alpha$-elements in these models are in reasonable agreement with the
values estimated by Brown, Wallerstein \& Oke (1990) and Brown \& Wallerstein
(1992). For this cluster, analysis of the CMD gives $\Delta V = 3.6 \pm 0.1$
(Hesser et al.\ 1987), which is the value that we have adopted for MRGC1 in
the above analysis.

As to the model parameters for MRGC2, we note that Zinn (1985) lists
${\rm [Fe/H]} \simeq -0.29$ for the MRGC NGC 6553, for which $\Delta V = 3.3$
(OBB90) or $\Delta V = 3.6$ (Ortolani 1994, based on analysis of improved
{\em HST} images). Thus, the corresponding models {\em may} be representative
of this cluster. As already emphasized, however, the actual abundances
of the $\alpha$-elements and helium in the cluster stars are, unfortunately,
still unknown in detail. Some insight may be gained from the available
estimates for {\em field} stars. For stars in the bulge, in particular,
Renzini (1994) has recently estimated, on the basis of the $R$-method,
a very high helium abundance: $Y \approx 0.30 - 0.35$. Minniti (1995a), on
the other hand, obtains a somewhat lower helium abundance for these stars,
$Y = 0.28 \pm 0.02$. McWilliam \& Rich (1994) find evidence that the field
stars in the bulge are overabundant in the $\alpha$-elements, although their
element ratios are not clearly compatible with the expected enrichment
patterns for SN II explosions. Also, very recently Idiart, de Freitas Pacheco
\& Costa (1995) obtained an integrated spectrum for the bulge in the direction
of Baade's Window. Their analysis of metallicity indicators like $Mg_2$ and
$\langle Fe\rangle$ points to an average metallicity ${\rm [Fe/H]} = 0.0$,
and an average ratio ${\rm [Mg/Fe]} = +0.4$. Comparison of these results with
the model predictions for the bulge (Fig. 5) would appear to favor Case 1.
Case 2 would not be ruled out either. Case 3, on the other hand, would appear
as an extreme possibility in applications to MRGCs that do in fact belong to
the Galactic bulge, since it implies oxygen abundances at high metallicities
which are similar to those obtained by Ratag et al.\ (1992) for the planetary
nebulae located in, but not necessarily belonging to, the Galactic bulge.
In this regard, it may be of interest as well to remark that very recently
Terndrup \& Walker (1994) have studied in detail the properties of NGC 6522
(located in Baade's Window, at $\simeq 1.5\, {\rm kpc}$ from the Galactic
center), finding this to be a metal-poor, blue-HB cluster --- similar to the
halo globulars located inside the solar circle. In fact, this leads us to
caution that the general correspondence between field and GC stars which are
located in a certain (limited) volume of space (in this case, the Galactic
bulge) must be established in more detail, employing (among others) the tools
from spectroscopic analysis.

Equally important, in this regard, is a precise identification of those
individual clusters that belong to the bulge and to the disk. Since these
two Galactic subsystems may have had different chemical enrichment
histories, it is clearly of the utmost importance to know the population
to which the cluster is associated in order to make a quantitative estimate
of its relative age in comparison with a reference object --- 47 Tuc, for
example. According to Zinn (1990, 1991), an individual ``bulge system of
GCs" probably does not exist. According to the recent analysis of a more
extended dataset by Minniti (1995b), on the other hand, several MRGCs may
indeed belong to the bulge. Our results show that special care should be
taken in comparing the ages of these objects with those for other clusters
lying outside the Galactic bulge, especially in what regards the assumed
chemical composition.

\section{On the expected location of the RGB bump}

In the present section, we present a short compilation and discussion of
recent theoretical and observational results for the RGB bump in GCs which
will hopefully prove useful in future comparisons with more accurate
observational data for the more metal-rich objects. The basic motivation
for our analysis are the suggestions by OBB90, Ortolani et al. (1992),
and Demarque \& Lee (1992) that the tilted HB clumps in some MRGCs
might be due to the proximity of the RGB bump.

The RGB bump occurs as an RGB star has a chemical composition profile
discontinuity, resulting from the previous inward penetration of its surface
convection layer, suddenly reached by the H-burning shell. When this happens,
the luminosity of the star drops, and it appears to hesitate momentarily prior
to continuing its journey to the tip of the RGB. As a result, a small, but
detectable (Iben 1968; King, Da Costa \& Demarque 1985; Castellani, Chieffi
\& Norci 1989; Fusi Pecci et al.\ 1990, hereinafter FFCRB90; Bono \&
Castellani 1992) agglomeration of stars is expected in the regions of the HR
and CM diagrams where this phenomenon takes place.

A comprehensive investigation of the occurrence of these RGB bumps in Galactic
globulars has been presented by FFCRB90. In this paper, the authors have
detected the bump in eleven Galactic globulars with metallicities in the range
$-2.15 \leq {\rm [Fe/H]} \leq -0.71$; the only previously reported detection
had been for 47 Tuc (King et al.\ 1985). FFCRB90 attempt to calibrate the
$M_{V}^{\rm HB}-{\rm [Fe/H]}$ relation on the basis of the magnitude
difference between the HB and the detected bump. They find, however,
zero-point problems in comparison with independent calibrations of this
relation, which demands an offset of the theoretical bump predictions by
some +0.4 mag in $M_{V}$ --- i.e., predicted bump locations are too bright
in comparison with the observations. Catelan (1992) calls attention toward
a secondary problem involving the metallicity dependence of their results.

While revised $M_{V}^{\rm HB}-{\rm [Fe/H]}$ relations have indeed implied
somewhat brighter RR Lyrae variables (see Catelan, de Freitas Pacheco \&
Horvath 1995 for a review and discussion of the evolutionary implications
of this result) than assumed by FFCRB90 in their analysis, a significant
zero-point offset is still present.

King et al.\ (1985) have argued that overshooting at the base of the
convective envelope is a likely explanation for this discrepancy, which they
as well had previously found. Similarly, Alongi et al.\ (1991) and Bono \&
Castellani (1992) note that existing bump models are based on the assumption
that there is a {\it sharp} discontinuity in the chemical composition profile
at the bottom of the convective layer, whereas in the real case a partially
mixed region is to be expected. From Table 1 in Alongi et al.\ 1991 and
Fig. 2 in Bono \& Castellani 1992, one may easily conclude that, upon
relaxation of this simplifying hypothesis, the bump can be made fainter by
a few tenths of a magnitude, while remaining a firm theoretical prediction.
A small contribution may also come from the effect of mass loss on the RGB
(Castellani \& Castellani 1993).

Improved opacities (FFCRB90) cannot alter the predicted bump magnitudes
significantly for low metallicities (Alongi et al.\ 1991; see also Ferraro
1992), although they may become important for higher metallicities (see
below). The influence of $\alpha$-elements, which may be important for the
bump magnitudes (Ferraro 1992), has been accounted for in the present paper
by computing $Z$ values from the Chieffi et al.\ (1991) expression.

More realistic, extensive models lacking, however, we are forced to restrict
the present analysis to the same theoretical framework adopted by FFCRB90.
We have employed Eqs. (2) and (6) in FFCRB90 for the magnitudes of the bump
and the ZAHB at $\log T_{{\rm eff}} = 3.85$. Recall that there is evidence
that Eq. (2) in FFCRB90 yields too bright bump magnitudes, so that the
$\Delta M_{V}$$^{{\rm HB}}_{{\rm bump}} \equiv M_V^{{\rm bump}}-M_V^{{\rm HB}}$
values thus derived should {\em not\/} be directly compared with observed ones.

The helium and $\alpha$-elements abundances have been drawn from the chemical
evolution models discussed in the previous section (cf. Table 1 and Fig. 5).

Values of $\delta\Delta V^{{\rm HB}}_{{\rm bump}}$ (where
$\delta \equiv {\rm MRGC1}-{\rm MRGC2}$) computed in this way can be found
in Table 2. The different entries correspond to different absolute ages
assumed for MRGC1 (column 1) and to different age ratios (column 2) between
MRGC1 and MRGC2, as obtained from Table 1.

Inspection of this table discloses that the bump in MRGCs with
${\rm [Fe/H]} \simeq -0.2$ should, according to the models, be located at a
fainter magnitude level (with respect to the cluster's HB) in comparison with
that in a MRGC like 47 Tuc, if Case 1 or Case 2 models are appropriate for
MRGC2. Case 3 models may imply somewhat brighter bumps in such clusters.
FFCRB90 determine, for 47 Tuc,
$\Delta V^{{\rm HB}}_{{\rm bump}} \simeq 0.45\pm 0.16$ mag; Ferraro et al.\
(1994) have found $\Delta V^{{\rm HB}}_{{\rm bump}} \simeq 0.52\pm 0.07$ mag
for the similarly metal-rich M69. These values may be used with the data in
Table 2 to estimate the actual expected location of the bump in MRGC2.
$V^{{\rm HB}}$ here refers to the lower envelope of the HB distribution.

The employed expressions for the magnitudes of the bump and the HB at the RR
Lyrae level have been obtained from models which employ old input physics ---
in
particular, old opacity tables. Both the Straniero \& Chieffi (1991) models,
for evolution prior to the HB, and the CCP91 models, for evolution on the HB,
represent a significant improvement in this respect, since they incorporate
more recent opacities in comparison with those upon which the FFCRB90 analysis
is based.

As a consequence of this difference in the input physics, the updated bump
and HB models become {\it progressively fainter} than the older models for
$Z > 10^{-3}$. They do agree quite well, however, at lower metallicities,
so that the conclusions by FFCRB90 for their metal-poorest clusters should
not be altered appreciably. For the metal-rich objects, on the other hand,
a possible correction term must be estimated.

This has been accomplished on the basis of Straniero \& Chieffi's (1991)
expression for the magnitude of the bump as a function of age and metallicity
and the present synthetic clump models based upon the CCP91 evolutionary
tracks.
Unfortunately, however, both studies are restricted to $Y_{{\rm MS}} = 0.23$,
so that we cannot analyze the direct impact of the new opacities upon the
results displayed in Table 2, which are based upon our chemical evolution
models. Also, the bump analysis of Straniero \& Chieffi is strictly valid
only for $Z \leq 6 \times 10^{-3}$, so that extrapolation toward higher
metallicities is needed.

Adopting $Z = 0.006$ for MRGC1 and $Z = 0.02$ for MRGC2, we find that improved
bump predictions for these objects become fainter than the earlier ones by
0.12 mag and 0.43 mag, respectively. This result is only weakly dependent on
the assumed age for the clusters. For the HB, on the other hand, the net change
amounts to $\approx 0.98-0.82 = 0.16$ mag and $\approx 1.33-0.86 = 0.47$ mag,
respectively. Since these values essentially cancel out in the computation
of $\Delta M_{V}$$^{{\rm HB}}_{{\rm bump}}$, the results given in Table 2
may be considered as actual standard {\em predictions\/} of the bump location
in MRGCs. If observed, deviations from these results should have important
implications for our understanding of the internal processes taking place in
red giant stars. The reader should bear in mind, however, that more extensive
models for several different chemical composition combinations are needed to
put these figures on a firmer basis.

A related question that may be raised concerning RGB bumps is: how numerous
are the bump stars expected to be in a MRGC? King et al.\ (1985), Castellani
et al.\ (1989) and FFCRB90 have all pointed out that the probability of
detection of the RGB bump increases with increasing metallicity, so that bump
stars may be relatively numerous in MRGCs.

To answer this question, one may make use of the theoretical relations which
have been obtained by Buzzoni et al.\ (1983). An illustrative example may be
given for the case $t_{9} = 10$, $Y_{{\rm MS}} = 0.28$, $Z = 0.02$. We apply
a correction factor to the bump lifetime estimate given by Buzzoni et al.\ in
their Eq. (5b), given that such an estimate apparently refers only to the phase
where the luminosity drops, whereas observationally what is of interest is the
full region on the RGB where the star crosses the same luminosity levels more
than once. A factor of 2.5 has been estimated, on the basis of a comparison of
the Bono \& Castellani (1992) and Castellani \& Castellani (1993) diagrams
with the predictions by Buzzoni et al.\ for the same chemical composition and
masses. Adopting an overall mass loss on the RGB $\Delta M = 0.15\, M_{\sun}$,
we find $t_{{\rm bump}}/t_{{\rm clump}} \approx 2.5\times 0.08 = 0.2$. This
result is essentially insensitive to the assumed amount of mass loss, given
that the HB lifetime is only weakly dependent on the stellar mass. The
following trends are present, though: decreasing the helium abundance by
$\Delta Y = 0.05$, we have $t_{{\rm bump}}/t_{{\rm clump}} \approx 0.31$;
decreasing the metallicity by $\Delta Z = 0.01$, we obtain
$t_{{\rm bump}}/t_{{\rm clump}} \approx 0.14$; increasing the age by
$\Delta t_9 = 5$, we find $t_{{\rm bump}}/t_{{\rm clump}} \approx 0.16$. To
summarize, roughly 20\% as many bump stars as clump objects should be present
in MRGCs, with the precise amount depending on the chemical composition and
age.
For comparison, a similar exercise assuming the Case 1 composition (cf. Table
1)
to be valid suggests
$t_{{\rm bump}}/t_{{\rm clump}} \approx 2.5\times 0.044 = 0.11$ for 47 Tuc,
for an age $t_9 = 14$. To be sure, these figures too may depend on the
assumptions about mixing at the bottom of the convective layer (cf. Fig. 2
in Bono \& Castellani 1992).

However, since RGB bumps have {\it not} been unequivocally identified in the
MRGCs observed by Ortolani, Barbuy \& Bica (1994 and references cited therein),
the comparison between the above predictions and the observations must be
postponed until the {\em HST} diagrams for these and other metal-rich objects
become available. The latter should represent an important test of our theories
of the evolution of red giants and the underlying input physics, and it is our
hope that the above summary will prove useful in future investigations of these
data.

\section{Summary and discussion}

One of the purposes of the present paper was to discuss the morphological
characteristics of the clump of stars in the CMDs of MRGCs which has been
associated with the core helium-burning phase (the HB). With this purpose,
synthetic HB models for metal-rich compositions, with and without differential
reddening, have been constructed and discussed.

We find that few combinations of chemical composition parameters and
evolutionary masses on the HB may lead to tilted synthetic HB models.
For high $Y$ values, some of the models do appear sloped in the $M_V,\, \bv$
diagram due to blanketing, and we expect that, for $Y_{\rm MS} > 0.3$, tilted
structures would become more common as a natural consequence of the properties
of the HB evolutionary tracks for high $Y$. In these models, the blueward loops
of stars evolving away from the red ZAHB are characterized by a slope
${\rm d}\log\,L/{\rm d}\log\,T_{{\rm eff}} > 0$ --- the slope increasing with
increasing $Y$. These trends could not be pursued any further in terms of
synthetic calculations in the present study, due to the lack of sufficiently
extensive grids of evolutionary tracks for high $Y$. We emphasize the need of
synthetic calculations employing {\em extensive grids} (in $Y$, $Z$, and
$M_{\rm HB}$) of HB evolutionary tracks to perform more detailed comparisons
with the observed CMDs, since neither ZAHB sequences nor individual
evolutionary tracks are expected to be able to adequately represent the
observed structures (Dorman et al.\ 1989).

The much more common ``clumpy" structures that we find may become tilted due
to the effect of differential reddening in the field of a MRGC, although the
computations suggest that a large amount of differential reddening should be
present in order to induce a noticeably sloped HB.

The age differences between two MRGCs are found to depend sensitively on
the chemical evolution scenario adopted for these objects, an important
point to bear in mind in future applications of the $\Delta V$ method,
especially since disk and bulge globulars may not have shared a common
chemical enrichment history. Indeed, disk and bulge globulars of similar
$\Delta V$ and [Fe/H] may not have the same ages, if their $Y$ and
[$\alpha$/Fe] are appreciably different. In particular, for a given
$\Delta V$ and [Fe/H], a cluster which was formed from material enriched
predominantly by SN II ejecta will appear {\em younger\/} than another one
which also suffered enrichment from SN Ia explosions. Depending on the
chemical evolution scenario adopted for the bulge, our analysis shows that
the bulge globulars may be younger or older than a cluster like 47 Tuc. We
believe that this should serve as an important motivation for observers to
determine the abundances of the $\alpha$-elements and of helium in MRGCs.

Zinn (1990, 1991) has argued that there may not be a genuine bulge system of
GCs in the Galaxy. Minniti's (1995b) more recent analysis, on the other hand,
suggests that at least some of the MRGCs may belong to the bulge. More
critical work appears needed to classify the {\em individual} MRGCs into disk
and bulge subsystems: our analysis demonstrates that the possibly different
chemical evolution histories of the disk and bulge globulars will have to be
seriously considered in more quantitative analyses of their CMD properties.

The predicted location of the RGB bump has also been critically discussed
in the present work. We provide magnitude estimates for the bump in a MRGC
with [Fe/H] $= -0.2$ for different chemical evolution scenarios. The RGB bump
is expected to be roughly 20\% as well populated as the helium-burning clump
in the same cluster.

\acknowledgments
The authors wish to thank Drs. S. Ortolani, B. Barbuy, and E. Bica for useful
discussions; they are especially grateful as well to Dr. B. Carney for his
careful reading of a draft of this paper, and for providing several relevant
comments and suggestions. M. C. would like to thank Dr. Ben Dorman for useful
discussions on the subject of model atmospheres computations, and J. C. Correia
and F. G. Garcia for their help in obtaining one key reference. Financial
support by FAPESP is acknowledged (grant 92/2747-8).

\clearpage

\begin{figure}
 \caption{ Synthetic clump models in the $M_{V},\, \bv$ diagram. The basic
 assumptions made in the calculations are summarized in the upper left portions
 of the panels. }
\end{figure}

\begin{figure}
 \caption{ In panel a, a clumpy synthetic HB model, similar to those depicted
 in Fig. 1, is displayed. In panels c through d, increasing amounts of
 (uniform) differential ``reddening" are assumed to be present in the field of
 the cluster:
 $\Delta E(\bv ) = 0.06, 0.10, \,\,{\rm and} \,\, 0.15\, {\rm mag}$,
 respectively. Besides increasing the dispersion in the diagrams, differential
 reddening clearly works in the sense of producing sloped structures: formal
 least-squares solutions with \bv as the independent variable give
 ${\rm d}V/{\rm d}(\bv ) \simeq 0.02, 0.33, 0.74,\,\, {\rm and}\,\, 1.26$
 for the models displayed in panels a through d, respectively. }
\end{figure}

\begin{figure}
 \caption{ As in Figs. 2a and b, except that a model already presenting a
 significant slope in the observational diagram {\it without} differential
 reddening having been included (panel a) is shown. The formal least-squares
 slope values in panels a and b are
 ${\rm d}V/{\rm d}(\bv ) \simeq 0.69\,\, {\rm and}\,\, 0.80$, respectively. }
\end{figure}

\begin{figure}
 \caption{ Synthetic clump model equivalent to Fig. 3a in the
 $M_{{\rm bol}} - \log T_{{\rm eff}}$ diagram. To be noted is that the
 blanketing effect is largely responsible for the production of a significant
 tilt in the CMD of Fig. 3a. }
\end{figure}

\begin{figure}
 \caption{ The helium abundance by mass $Y$ (panel a) and oxygen-to-iron ratio
 [O/Fe] (panel b) as a function of the metallicity [Fe/H], as predicted by
 three different models for the chemical evolution of the Galactic bulge. See
 text for more details. }
\end{figure}

\end{document}